\documentclass[a4paper]{jpconf}
\usepackage{graphicx}
\usepackage[colorlinks=false]{hyperref}
\usepackage{xcolor}
\usepackage{float}					   

\begin{document}
\title{Removeing Noise From Simulated Events at The Main Drift Chamber of BESIII Using Convolutional Neural Networks}

\author{Hosein Karimi Khozani, Zhang Yao and Yuan Ye}

\address{Institute of High Energy Physics, Shijingshan, Beijing, China}

\ead{karimi@ihep.ac.cn}

\begin{abstract}
	BESIII is the particle detector of the Beijing Electron-Positron Collider, which is a 
    $\tau$-charm factory working at 
    energies around 4 GeV. The first part of the detector, around the collision site, 
    is called the Main Drift Chamber, MDC. The events recorded at MDC are 
    mixed with the background noise of various origins. On average, about 10\% of the hits 
    of an event are noises. Still, the noise level differs event by event, and 
    some of the events might even get 
    more noise hits than signal hits, making the analysis less efficient. 
    The standard algorithms of the offline software system of BESIII reconstruct 
    signal tracks using the polluted data. This reduces the reconstruction efficiency 
    of high noise tracks. In this article, we test the idea of using supervised deep learning techniques 
    to remove this noise beforehand. We generate Monte Carlo events, then mix them with
    noise hits coming from real data. At first, we use deep learning techniques to 
    classify the hits based on their individual features. Then, we simplify every event to a 40 by 43 picture and use 
    image recognition tools to remove the noise. The average noise level for these events 
    with only two signal tracks is about 30\%. 
    On average, the techniques presented in this article can purify Bhabha events 
    to nearly 99\% while preserving about 99\% of the signal tracks.
\end{abstract}

\section{Introduction}
MDC is a 2.6m long cylindrical drift chamber with an outer diameter of 1.6m. It has 6796 sense wires 
which are distributed in 43 layers. There are two levels of trigger 
filtration in BESIII before the data gets stored. The 
efficiency of this system is as high as 100\% for Bhabha 
scattering, and 97.7\% for any $J/\Psi$ decay mode \cite{BESIII:2009fln}. At the same time, the background 
rate reduces significantly, from $10^4$ KHz to about 1 KHz out of 3 KHz, which is the rate 
of event storage on the offline system \cite{BESIII:2009fln}. 
This is an efficient process, but still, one third of this data is uninteresting background. 
Furthermore, the number of hits of an event is typically between 100 to 200, which about 
5 to 15 percent of them are noises. The 
standard packages used on the offline system should remove the rest of the 
noise. Basically, these packages find the tracks of interest by considering all the hits, 
and the noise is eventually left aside like background tracks. Therefore it is hard 
to analyze events with high levels of noise. Our motivation is to reduce 
noise beforehand with deep learning methods while keeping efficiency high.

We present the first phase of our project here, in which we worked with simulated Bhabha 
events. Working with these simple events is a suitable choice for starting this project. Moreover, BESIII events 
are, in general, very clean. Though the luminosity is high, still most of the events 
have only two or four signal tracks. Another notable point is that Bhabha events are 
useful for purposes such as accurate luminosity measurements. Initially, we 
look at the situation as a hit classification problem and form neural networks that can 
distinguish noise hits solely based on their features. The features, in general, include 
time and charge, but we here only work with raw-time values. For the next step, we form 
convolutional networks to look at the whole event at once and remove the noise. These 
models basically remove the noise by finding the tracks, just like the standard algorithms. 
We later continued this work with graph neural networks and with more complicated events. 
The results will show up in an upcoming paper soon. The lesson from those studies is 
that the networks we find here need little change for more complicated situations and 
are indeed applicable to them as well. 

\section{Data Generation}
MDC maps the 3D image of the events to 2D by recording the hits, which show up as electric 
pulses at the end caps. Still, the whole 3D image of an event is reconstructable, thanks 
to the time of 
flight information and the fact that the sense wires are not all aligned. The wires in 
the \textit{straight}
layers are aligned with the $z$ axes. However, wires of the \textit{stereo} layers have angles 
with this direction. 

As mentioned above, we work with Bhabha events for the start. These scattering events 
only include two tracks of high energy electrons and positrons. The solenoid magnetic field 
at MDC is about 1 T. The diameter of the detector is about 1.6 m. Naturally, the tracks 
are straight lines. Having 43 layers means there are around 86 signal hits for each 
event. \verb"Bhwide" generator is used to simulate these events \cite{Ping:2008zz}. 
We do not work with all the available 
information, which would be accessible for real data only after the reconstruction process. 
Instead, only what would be in the outcome of MDC is considered, which includes two 
geometrical dimensions plus the time that the hit happens and its electric charge value.

Next, 
we add noise to these events using another package of the offline system of the 
spectrometer called \verb"BESEventMixer" \cite{Ping:2008zz}. This algorithm uses real 
data and randomly adds noise of various 
origins to the simulated events. The noise hits have only space and time related features. 
The added noise 
is therefore independent of the events as expected and is, on average, about 40 hits per 
event. Hence, about one third of our data is noise, and the neural networks that we form 
have the task of removing it. We simulated 80000 events, which was enough 
for our purposes here.

\section{Fully Connected Networks}
\begin{figure}[h]
    \begin{center}
    \begin{minipage}{14pc}
    \includegraphics[width=14pc]{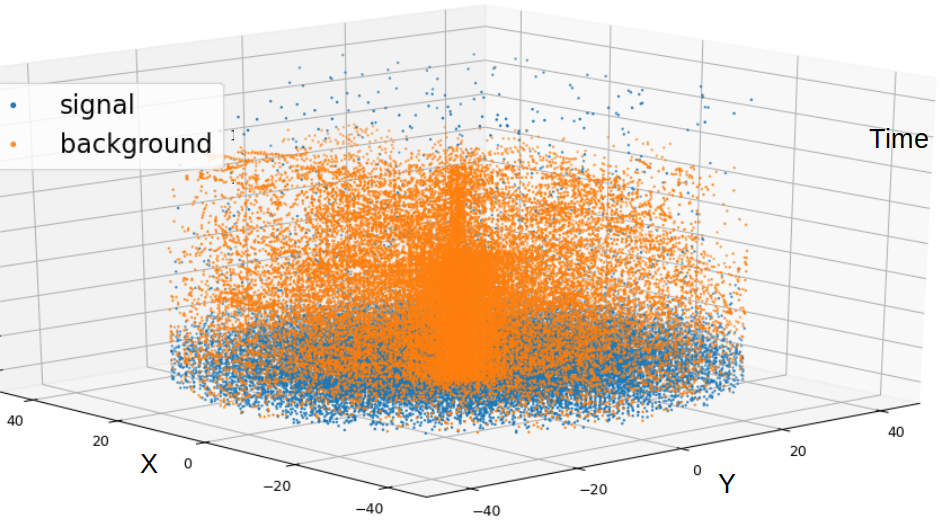}
    \caption{\label{featurespace}Feature space for the hits of about 400 
        Bhabha events. Noise hits (orange dots) happen uniformy over time.}
    \end{minipage}\hspace{2pc}
    \begin{minipage}{14pc}
    \includegraphics[width=14pc]{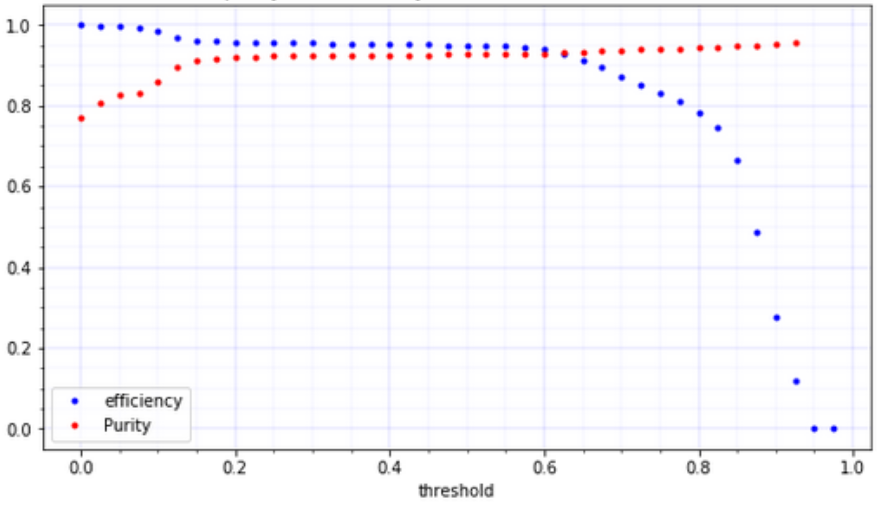}
    \caption{\label{purefffcn}Purity and efficency of the results of 
        the fully connected network versus threshold of probablity.}
    \end{minipage}
    \end{center}
\end{figure}

For the first step, we choose the regression approach to classify the hits  
into signal or background. 
The feature space has two space dimensions and one time dimension [Fig. \ref{featurespace}]. 
We developed a neural 
network that is, in principle, able to learn a threshold in time for every cell of every 
layer that discriminates between signal and background. This is related to the time that 
the generated particle most likely reaches that point. Such a network cannot make use of the geometry of the 
trajectories, which is the motivation behind using convolutional networks introduced in 
the next section. In fact, all the hits are introduced to the network simultaneously and not 
as in individual events. Nonetheless, the outcome is better than a simple cut line in time for a few 
percents.
    
The model which works the best has seven fully connected hidden layers. This 
network can increase the original 70\% average purity of the events to 93\% and keep 
94\% of the signals [Fig. \ref{purefffcn}]. In other words, it removes 85\% of the noise 
with the expense of losing about 6\% of the signal hits. This result is the baseline 
for us to compare the effectiveness of other models, such as 
the CNN model introduced in the next section and the graph neural networks which will be 
presented in another article. 

Since this model sees the hits one by one, we can apply it to one 
part of the data if needed. So it can be used in combination with other models. For example, 
in the next section, we show how to simplify the events to pictures with fewer resolutions. 
If we apply this dense model to the data lost during simplification, then 
the efficiency of the whole model is comparable to when no simplifications are involved, 
while it is a faster and lighter model in terms of memory use.

\section{Convolutional neural networks}
The 3D trajectory of particles in MDC lit the drift cells, which send their signals to 
the end caps at the ends of their lengths. Thus, our data in its raw format 
includes the 2D projection of the 3D events. The pixels of each
image are arranged in 43 concentric circles, each with a different resolution. The number 
of cells gradually increases from 40 on the 
first two layers to 288 on the last three. 
There are several acceptable ways to map this 
to a square to feed the neural networks. Here, we choose a simple method of 
looking at the events with the resolution of the first layer. So, we map every event to 
a 2D image with a size of 43 by 40. This means that the information 
of some of the neighboring cells is combined at the outer layers. 

\begin{figure}[h]
    \begin{center}
    \begin{minipage}{24pc}
    \includegraphics[width=13pc]{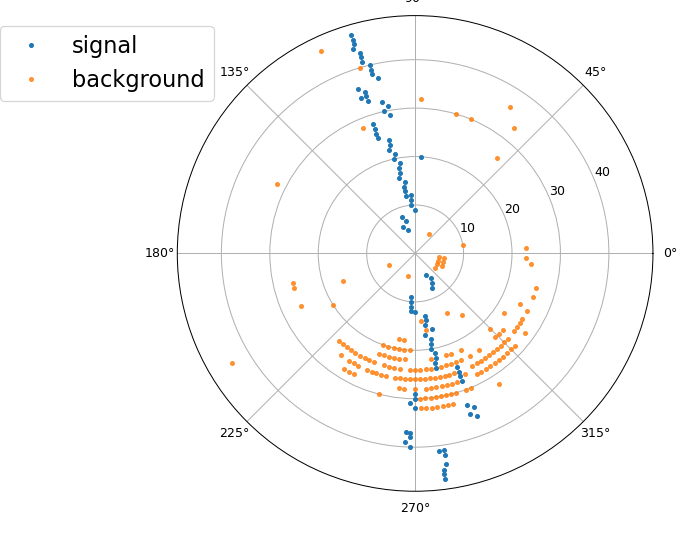}
    \includegraphics[width=11pc]{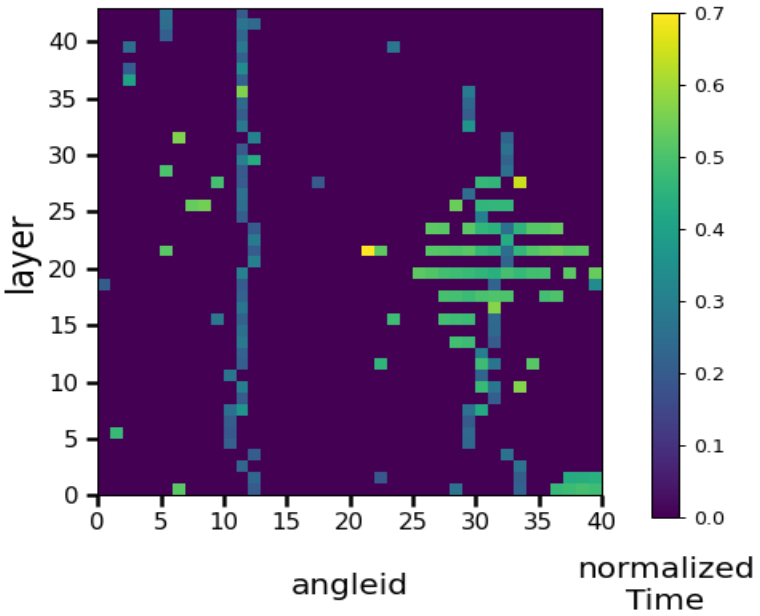}
    \caption{\label{eventcnnim} A sample event recorded at MDC is shown on the left 
    side. The right side image shows the same event mapped on a 40 by 43 pixels image.}
    \end{minipage}\hspace{2pc}
\end{center}
\end{figure}

With this simplification, the neural network cannot make use of the information of 
about 2\% of the hits. When that happens, 
one of the things that we can do is to just keep 
the hits that happened earlier since they are more likely to be signal hits. A better 
way is to pass that part of the hits from the classifying network introduced in the 
previous section. It was mentioned there that just 
looking at the time and position values of the hits can lead to more than 85\% noise 
deletion and 
about 94\% signal preservation. If we apply this to the 2\%, this simplification of 
the images removes about 0.1\% of the signal hits and pollutes the data for only about 
0.2\%. The real situation is even better since the hit classification results are more 
efficient better for the hits happening at the last layers. Therefore we take this simplification of the images 
as a good approximation. Furthermore, we normalize the time value of each hit and show 
it to the network just like color is presented for normal pictures (Fig. \ref{eventcnnim}).

\begin{figure}[h]
    \begin{minipage}{36pc}
    \includegraphics[width=12pc]{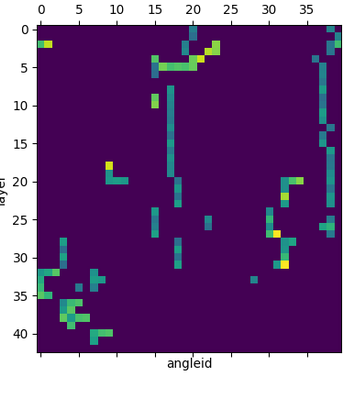}
    \includegraphics[width=12pc]{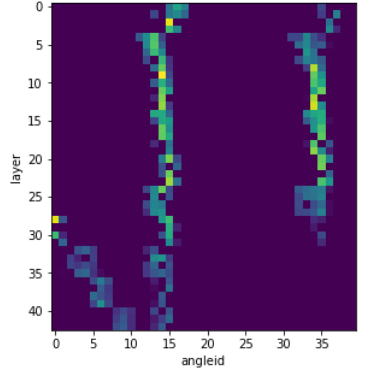}
    \includegraphics[width=12pc]{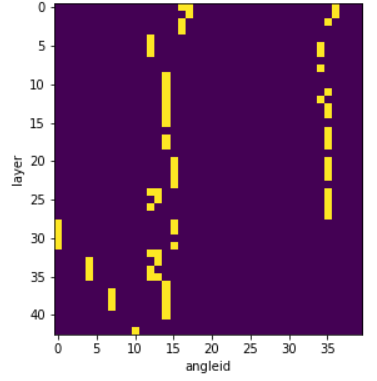}
    \caption{\label{thresholdim} A random event which includes noise is shown 
    on the left side. The image at the middle shows the outcome of the network for 
    this image. The right side image shows the same result after applying a threshold.}
    \end{minipage}
\end{figure}

Our investigations show that a deep learning model with five convolutional layers and five  
dense layers can successfully remove noise hits while preserving signal tracks. This 
network does the job by detecting the tracks along with the use of time values. 
This situation is just like when the color of the pixels helps the segmentation process 
for regular images. We observe that it is necessary to downsize mildly with the filtering 
process of 
the convolutional layers, then flatten them and pass the results through dense layers. The last hidden layer has 64 
nodes. The target layer has 1720 nodes which is the number of pixels needed to 
rebuild the pictures. It is not needed to have convolutional layers for upsampling 
which are present in models like Unet \cite{unet}. In the 
final image, the particle tracks are boldened, and the off-track noise is gone perfectly. 
Moreover, we test different thresholds to keep the hits. This way, the tracks become 
sharper, and the on-track noise also disappears. In the end, we intersect the result with 
the input to not have any extra hits. Figure  
\ref{thresholdim} show the result of passing an event from the trained network 
and then applying the threshold.

\begin{figure}[h]
    \begin{center}
    \includegraphics[width=20pc]{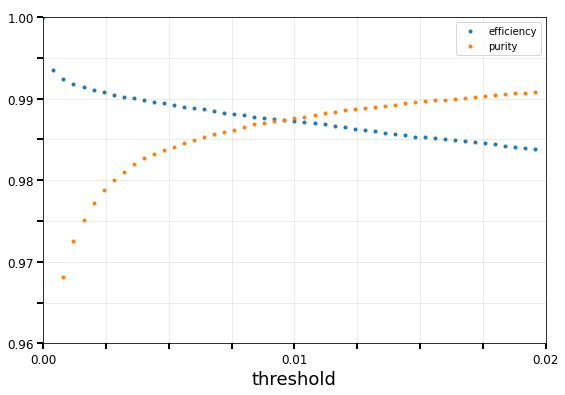}
    \end{center}
    \begin{center}
    \begin{minipage}[b]{30pc}\caption{\label{cnnresult}This diagram shows the average 
        efficiency of the signal preservation and the average purity of the test events 
        after they pass through the trained network.}
    \end{minipage}
    \end{center}
\end{figure}

We train the network with 77000 events up to about 100 epochs using the IHEP GPU farm. 
The result is an increase in the 
average purity of the events from the original amount of about 70\% to 98\% while  
about 99\% of the signal hits are preserved (Fig. \ref{cnnresult}). This means that, on average, 
every event will have only about 1.5 noise hits and lose only one signal hit after 
passing through the trained network.

We examined many possibilities and came up with the above architecture of the 
convolutional network, which is a 
rich model with about 20 million trainable parameters. It is worth mentioning that for 
different purposes, different areas of Fig. \ref{cnnresult} might be favorable. 
We usually use the 
crossing point of purity and efficiency diagrams to compare different models. After 
this work, we 
tried more complicated events and more detailed images, and it turned out that the same 
structure worked for them as well. As for the Bhabha events, we try 100 by 100 pixel 
images which have comparable resolution with the real 2D images, e.g. 6796 pixels. 
They also lead to similar results and show that the above model is acceptable. We will 
discuss them along with more techniques and results in an upcoming paper. 

\section{Concolusion}
Since we already have many tools in high energy physics to generate simulated data and 
real data has also been stored for many years, it is very reasonable to use machine 
learning techniques in this area. Moreover, even a slight improvement in the purification 
of the stored events 
can be important in physics analysis. We see that even simplified models can have acceptable 
performances while being very fast and light which makes them good candidates for the    trigger 
system  
or pre-processing of data. The hit classification approach for Bhabha events at the 
drift chamber of BESIII 
detector makes a baseline for further investigations. It also shows that a regression 
approach can be used in combination with other methods. Finally, the convolutional neural network 
that we introduced can distinguish the particle tracks and remove the noise in them 
in a way that the efficiency of the process and the purity of the final events are both 
about 98.7\% at the same time.

\section*{References}
\bibliography{ref-cnnnoiseremoval}{}
\bibliographystyle{plain}
\end{document}